# Implicit Government Guarantee Measurement Based on PMC Index Model

Yan Zhang, Yixiang Tian, Lin Chen, Qi Wang
*School of Management and Economics, University of Electronic Science and Technology of China, Chengdu, China*
Corresponding author: chenlin2@uestc.edu.cn

**Abstract**: The implicit government guarantee hampers the recognition and management of risks by all stakeholders in the bond market, and it has led to excessive debt for local governments or state-owned enterprises. To prevent the risk of local government debt defaults and reduce investors' expectations of implicit government guarantees, various regulatory departments have issued a series of policy documents related to municipal investment bonds. By employing text mining techniques on policy documents related to municipal investment bond, and utilizing the PMC index model to assess the effectiveness of policy documents. This paper proposes a novel method for quantifying the intensity of implicit governmental guarantees based on PMC index model. The intensity of implicit governmental guarantees is inversely correlated with the PMC index of policies aimed at de-implicitizing governmental guarantees. Then as these policies become more effective, the intensity of implicit governmental guarantees diminishes correspondingly. These findings indicate that recent policies related to municipal investment bond have indeed succeeded in reducing implicit governmental guarantee intensity, and these policies have achieved the goal of risk management. Furthermore, it was showed that the intensity of implicit governmental guarantee affected by diverse aspects of these policies such as effectiveness, clarity, and specificity, as well as incentive and assurance mechanisms.

## 1. Introduction

Many cities in China have a state-owned enterprise called the city development investment company, which has issued many bonds called municipal investment bonds. For example, 4.89 trillion RMB of municipal investment bonds were issued in 2023.The fundamental property of municipal investment bonds is belonged to the corporate bonds. However, since the city development investment company is a state-owned enterprise, Funds raised through municipal investment bond financing are primarily used for infrastructure investment in local cities and urban development. Although the repayment funds for municipal investment bonds are mainly derived from the company's profits, the urban infrastructure investment projects have a long cycle and low returns, and the repayment funds often rely on local government revenue. Therefore, it is generally believed that municipal investment bonds are supported by the credit of the local government and are often referred to as "quasi-municipal bonds". However, this guarantee is not a clearly stated legal requirement and is referred to as an implied government guarantee. Because the city investment company is a local state-owned enterprise, when the city investment company may default, the local government, in order to meet its ongoing financing needs, will provide support for the repayment of the city investment company's debt. This serves as an illustration of an implicit government guarantee [1-2].

Although local governments do not have an obligation to guarantee bonds issued by the city investment companies, but Luo and Liu (2020) revealed that there was no statistically significant difference in the credit spread between unsecured municipal investment bonds and third-party guaranteed municipal investment bonds, it is suggested that the market perceives implicit government guarantees associated with unsecured municipal investment bonds [3]. Because if the municipal investment bonds default, which would significantly undermine the credit rating and future financing capabilities of city development investment company. Consequently, the local government is likely to extend support from various avenues to its municipal investment platform in an effort to facilitate the repayment of bonds under its jurisdiction（Li，2020）. Moreover, research indicates that implicit government guarantees exert varying effects in regions characterized by favorable fiscal conditions compared to those with less favorable fiscal circumstances [4-5]. In summary, analyzing the implicit government guarantee and regulatory policies of China's municipal investment bonds can have significant implications for the municipal platform's ability to prevent debt crises and promote marketization [6-7].

There are generally three methods for measuring government implicit guarantees: the first method mainly uses indicators such as local government financial revenue and local economic development speed as indicators to measure local government implicit guarantees. This method may have significant heterogeneity due to the different degrees of government implicit guarantees between bonds [8]. The second method uses the degree of government support for state-owned enterprises (such as interest rate incentives, government subsidies, loan support, etc.) to measure implicit government guarantees, but these data mainly reflect explicit guarantees. The third method is relatively innovative, such as the Merton option pricing formula and orthogonal decomposition method. However, due to many state-owned enterprises in China being non listed companies, it is difficult to calculate the parameters required for the Merton option pricing formula, and the application of this model is limited.

The intrinsic property of the implicit government guarantee is a kind of expectation. Diminishing the reliance on this implicit government guarantee for municipal investment bonds can facilitate a balance between financial risk mitigation and local government debt management, thereby fostering high-quality development in both finance and the economy [9]. In response, the central government and pertinent regulatory bodies have released a series of policy documents aimed at governing the development of municipal investment bonds. For example，In September 2014, the State Council issued a document named 《Opinions on Strengthening the Management of Local Government Debt》,in order to limit local governments from borrowing through enterprises. But Choi and Lu et.al (2022) find evidence that No. 43 Document, which was introduced to curb the "incorrect" investor perception of implicit guarantees on MCBs, has had little effect in that its introduction has not significantly changed. On October 27, 2016, the General Office of the State Council issued 《a notice on the Emergency Disposal Plan for Local Government Debt Risks》. Have these policies changed the implicit government guarantee expectation for municipal investment bonds? Have they had a risk management effect? Some scholars have studied specific documents, but there has been no comprehensive systematic research.

In summary, this paper employs advanced computer text analysis techniques to analysis 17 pivotal policy documents pertaining to the standardized development of municipal investment bonds. Furthermore, it utilizes the policy modelling consistency (PMC) model to assess the efficacy of these policies in mitigating implicit government guarantees. Subsequently, a framework for quantifying the degree of implicit guarantees associated with municipal investment bonds is developed. This paper makes two primary contributions. First, it assesses the effectiveness of de-implicit guarantee policies for municipal investment bonds through the lens of the PMC index model. The PMC index model is a comprehensive evaluation method used to assess the effectiveness of policies [11-12]. Second, it introduces a novel methodology for quantifying the strength of implicit government guarantees. This approach offers fresh insights and tools for investors and regulatory authorities to identify and evaluate risks associated with municipal investment bonds, thereby enhancing their understanding of these risk characteristics, and fostering the healthy development of the market.

The remainder of this paper is organized as follows: Section 2 introduces a measurement methodology for government implicit guarantees utilizing the PMC index model; Section 3 presents a textual analysis of policy documents pertinent to municipal investment bonds; Section 4 encompasses both the computation of the PMC index and an assessment of government implicit guarantee intensity derived from this textual analysis; Section 5 summarizes the conclusions.

## 2. Implicit government guarantee measurement based on PMC index

The process of constructing the implicit government guarantee measurement model based on the PMC index is as follows.

First, it is the keyword extraction, which entails identifying pertinent keywords from policy documents associated with the municipal investment bonds. This process is a vital component of financial policy analysis, significantly enhancing the efficiency of recognizing and comprehending policy information. The methodology primarily uses structured processing of textual data to thoroughly investigate the themes within policy texts and extract critical information. To facilitate an in-depth exploration of thematic terms related to government implicit guarantees within various management documents concerning municipal investment bonds, this study encompasses all relevant policy texts in its keyword analysis. By synthesizing results derived from both the TF-IDF algorithm and Text-Rank algorithm, this research aims to distill thematic words that encapsulate the core of government implicit guarantee policies for municipal investment bonds.

Second, it is co-word analysis. This process entails a statistical examination of the frequency with which keywords appear across various documents, thereby illustrating the degree of association among these keywords. The relationships between keywords form a co-word network, where both the proximity and connectivity of keywords within this network offer a direct visual representation of the interrelations among the topics each keyword signifies. In this paper, we use hierarchical clustering analysis to categorize the extracted thematic words, aiming to further elucidate the intrinsic characteristics of implicit government guarantee policies concerning city investment bonds.

Finally, the strength of implicit government guarantee is computed based on PMC index.

The PMC index model is particularly well-suited for the analysis and evaluation of complex systems associated with economic and social policies. The series of policy documents concerning municipal investment bonds issued by government aims to mitigate public expectations regarding implicit government guarantees while enhancing the risk management capabilities of all stakeholders involved. Should these policies fulfill their intended objectives, we would anticipate an increase in the PMC index alongside a corresponding reduction in implicit government guarantees. We propose a straight forward yet robust conversion formula linking the PMC index to implicit government guarantees, thereby introducing a novel methodology for calculating such guarantees.

In the next section, we will provide a detailed introduction to the index system for evaluating the policy effects of standardizing the development of municipal investment bonds.

## 2.1 Indicators variables of PMC Index

In order to construct the PMC index to evaluate the consistency of policy documents of municipal investment bonds, we first extensively collected policy documents related to municipal investment bonds and selected the sample of policy documents from various national-level departments and agencies, including the State Council, the Ministry of Finance, the People's Bank of China, and the National Development and Reform Commission. Based on the studies of existing PMC index models, a set of indicators was established for evaluating the policy documents on standardizing the development of municipal bond issuance [10].

（1） Main-variables and sub-variables of PMC-Index

The construction of the PMC-Index involves the use of forty-two (42) sub-variables distributed in ten (10) main-variables. These 10 main-variables are: characteristics of policy ($P_1$), validity of policy ($P_2$), areas of policy application ($P_3$), policy sources ($P_4$), incentives and guarantees of policy ($P_5$), functional of policy ($P_6$), operational levels of policy ($P_7$), subjects addressed of policy ($P_8$), impact levels of policy ($P_9$), and transparency of policy ($P_{10}$). $P_1$、$P_2$、$P_4$、$P_9$、$P_{10}$ evaluate the quality and transparency of policies from the perspective of their fundamental attributes. $P_3$、$P_5$、$P_6$、$P_7$、$P_8$ from the perspectives of policy coverage, incentive mechanisms, function positioning, influence scope, and target objects, the paper identifies the policy of municipal investment bonds, in terms of their ability to manage risks, address the problem of implicit guarantees, and regulate and adjust. The main-variables and sub-variables are showed in Table 1.

Table 1．main-variables and sub-variables for the PMC Index

| main-variables | | sub-variables | |
|---|---|---|---|
| $P_1$ | characteristics of policy | $P_{11}$ | Prediction |
| | | $P_{12}$ | Regulation |
| | | $P_{13}$ | Recommendation |
| | | $P_{14}$ | Describe |
| | | $P_{15}$ | Identification |
| | | $P_{16}$ | Orientation |
| $P_2$ | validity of policy | $P_{21}$ | Long-term |
| | | $P_{22}$ | Midterm |
| | | $P_{23}$ | Short-term |
| $P_3$ | areas of policy application | $P_{31}$ | Economy |
| | | $P_{32}$ | Society |
| | | $P_{33}$ | Technology |
| | | $P_{34}$ | Politics |
| | | $P_{35}$ | Environment |
| $P_4$ | policy sources | $P_{41}$ | State Council |
| | | $P_{42}$ | Government ministries |
| | | $P_{43}$ | Provincial and municipal party committees |
| | | $P_{44}$ | Provincial and municipal departments and bureaus |
| $P_5$ | incentives and guarantees of policy | $P_{51}$ | Legal safeguards |
| | | $P_{52}$ | Technical guidance |
| | | $P_{53}$ | Financial support |
| | | $P_{54}$ | Tax reduction and exemption |
| | | $P_{55}$ | Investment subsidy |
| $P_6$ | functional of policy | $P_{61}$ | Standard guidance |
| | | $P_{62}$ | Risk prevention |
| | | $P_{63}$ | Investments Optimizing |
| | | $P_{64}$ | Infrastructure |
| | | $P_{65}$ | Establish a robust system. |
| $P_7$ | operational levels of policy | $P_{71}$ | National Development |
| | | $P_{72}$ | Regional Economy |
| | | $P_{73}$ | Industrial structure |

|   |   | $P_{74}$ | Business operation |
|---|---|---|---|
|   |   | $P_{75}$ | Product Standards |
| $P_8$ | subjects addressed of policy | $P_{81}$ | local government |
|   |   | $P_{82}$ | Business |
|   |   | $P_{83}$ | Financial institutions |
|   |   | $P_{84}$ | The general public |
| $P_9$ | impact levels of policy | $P_{91}$ | Laws and regulations |
|   |   | $P_{92}$ | Administrative Regulations |
|   |   | $P_{93}$ | Department regulations |
|   |   | $P_{94}$ | Standards documents |
|   |   | $P_{95}$ | Industry Standards |
| $P_{10}$ | transparency of policy | $P_{10}$ |   |

（2）The use of multi-input-output table

The multi-input-output table is an alternative database analysis framework that permits storage of a large amount of data to measure any single variable (see Table 2). the multi-input-output table functions as the basic analytical framework to measure the "$n$" number of main-variables. Each main-variable is formed by "$m$" number of sub-variables. The number of sub-variables in each main-variable is unlimited. As such, the multi-input-output table concept does not include any notion of ranking of variables according to importance(Estrada，2011）.

The main-variables encompass various dimensions, including policy nature, policy effectiveness, and policy domain. The sub-variables further delineate these dimensions by providing more specific evaluative indicators. Each sub-variable associated with the main-variable has been meticulously crafted to ensure a comprehensive representation of all facets of the policy. All of 42 sub-variables are given the same importance (weight) because we are interested to measure a single value, which is the PMC-Index in this case.

Table 2. Multi-input-output table

| Main-variables | Sub-variables |
|---|---|
| $P_1$ | $P_{11}$:1  $P_{12}$:2  $P_{13}$:3  $P_{14}$:4  $P_{15}$:5  $P_{16}$:6 |
| $P_2$ | $P_{21}$:1  $P_{22}$:2  $P_{23}$:3 |
| $P_3$ | $P_{31}$:1  $P_{32}$:2  $P_{33}$:3  $P_{34}$:4  $P_{35}$:5 |
| $P_4$ | $P_{41}$:1  $P_{42}$:2  $P_{43}$:3  $P_{44}$:4  $P_{45}$:5 |

| $P_5$ | $P_{51}$:1 $P_{52}$:2 $P_{53}$:3 $P_{54}$:4 $P_{55}$:5 |
| $P_6$ | $P_{61}$:1 $P_{62}$:2 $P_{63}$:3 $P_{64}$:4 $P_{65}$:5 |
| $P_7$ | $P_{71}$:1 $P_{72}$:2 $P_{73}$:3 $P_{74}$:4 $P_{75}$:5 |
| $P_8$ | $P_{81}$:1 $P_{82}$:2 $P_{83}$:3 $P_{84}$:4 $P_{85}$:5 |
| $P_9$ | $P_{91}$:1 $P_{92}$:2 $P_{93}$:3 $P_{94}$:4 $P_{95}$:5 |
| $P_{10}$ | $P_{10}$:1 |

The establishment of the multi-input-output table not only provides a structured analytical framework for policy evaluation, but also lays a foundation for subsequent quantitative analysis. Through Table 2, policy indicators can be systematically organized and quantified, enabling the calculation of the PMC index for the risk reduction and guarantee of municipal bond management policy.

## 2.2 Measuring Government Implicit Guarantees Based on the PMC Index

The measurement of the government implicit guarantees based on PMC index involves five steps. (1) The first step is to put the 10 main variables and 42 sub-variables into the multi-input-output table. (2) The second step is to evaluate sub-variable by sub-variable according to the parameters mentioned above (see Eq (1) and (2)).

$$P_{ij} = \begin{cases} 1, & \text{The feature is explicitly showed in the policy text.} \\ 0, & \text{The feature is not explicitly showed in the policy text.} \end{cases} \quad (1)$$

$P_{ij}$ is the $j$ sub-variables of $i$ main-variables.

(3) The third step is to calculate the value of each main-variable. This value is the sum of all sub-variables divided by the total number of sub-variables (see Expression (3)).

$$P_i = \frac{\sum_{j=1}^{n_i} P_{ij}}{n_i} \quad (2)$$

$n_i$ is the number of sub-variables.

(4) The fourth step is the actual measurement of the PMC.

$$PMC = \sum_{i=1}^{10} \sum_{j=1}^{n_i} \frac{P_{ij}}{n_i} \quad (3)$$

(5) The last step is the actual measurement of the intensity of the implicit government guarantees. The calculation of intensity of the implicit government guarantee is presented in formula (4). A series of policies issued by various levels of government and departments aimed at managing municipal investment bonds are intended to eliminate investors' expectations of implicit government guarantees. Therefore, if these policies effectively reduce the implicit guarantee expectations and lower the risks of municipal investment bonds, the potential demand for implicit government guarantees will be lower, and the implicit government guarantee intensity will accordingly decrease. Therefore, based on the municipal investment bond policy, the government implicit guarantee intensity index is defined as follows, referring to the distribution of the PMC index.

$$G = 10 - PMC \qquad (4)$$

In above Eq (4), $G$ is a measurement of the intensity of government implicit guarantee intensity. According to the definition of variables, the PMC-Index is within the range of 0-10, so the intensity of the implicit government guarantees (G) is also within the range of 0-10. Table 3 provides the meaning of intensity of implicit government guarantee at the four levels.

Table 3 Four levels of implicit government guarantee

| Intensity of implicit government guarantee measurement | Evaluation for implicit government guarantee |
| --- | --- |
| [5,10) | Perfect implicit guarantee |
| [3,5) | Good implicit guarantee |
| [1,3) | Acceptable implicit guarantee |
| [0,1) | Low implicit guarantee |

Finally, the construction of the PMC-Surface. The purpose of constructing the PMC-Surface is to graphically represent all results in the PMC-Matrix. The PMC-Surface shows the strengths and weaknesses within any policy modeling on a multi-dimensional coordinate space. The construction of the PMC-Surface is based on the PMC-Matrix results. The PMC-Matrix is a three-by-three matrix that contains the individual results of all nine variables

$$PMC\ Surface = \begin{bmatrix} P_1 & P_2 & P_3 \\ P_4 & P_5 & P_6 \\ P_7 & P_8 & P_9 \end{bmatrix} \qquad (5)$$

# 3 Textual Analysis of Policy Documents

## 3.1 Samples of policy documents

We selected the most relevant 17 documents issued by various government departments from 2008 to 2024 related to the development of municipal investment bonds. The detailed information of the 17 policy documents is presented in Table 4.

Table 4. Samples of policy documents

| Number | File name of Policy | Release date | The goal of the policy |
|---|---|---|---|
| 1 | 《Notice on Related Matters Concerning the Development of Corporate Bond Market and Simplification of Approval Procedures》 | 2008.01 | Simplify the bond issuance process to facilitate greater role of borrowing financing in economic development. |
| 2 | 《Guidelines for Further Strengthening Credit Structure Adjustment and Promoting a Stable and Rapid Economic Development》 | 2009.03 | Propose supporting conditional local government financing platforms by issuing bonds to expand their financing channels. |
| 3 | 《Notice on Strengthening the Management of Local Government Financing Platform Companies》 | 2010.06 | Open the supervision of China's municipal government bond issuance, requiring further cleanup and handling of debts held by municipal investment companies, and strengthening financing supervision and credit supervision over municipal investment companies and financial institutions that lend to them. |
| 4 | 《Notice of the China Banking Regulatory Commission on Effectively Managing the Risk of Government-funded Project Loans for Local Governments in 2011》 | 2011.03 | Continue to implement the requirements of National Development Document No. 19, with the goal of "reducing old debts and controlling new ones," centralize the approval and management of platform companies according to a list, and carry out comprehensive management of city investment platform companies and platform loans in banks nationwide. |
| 5 | 《Notice on Stopping Illegal and Inappropriate Financing Practices by Local Governments》 | 2012.12 | It is required that local governments step up their management of government debt and prohibit them from providing illegal guarantees and commitments for non-government debt. |
| 6 | 《Notice on the Use of Corporate Bond Financing to Support the Renovation of Shanty Areas》 | 2013.08 | Support for national key projects. |

| 7 | 《Opinions on Strengthening the Management of Local Government Debt》 | 2014.10 | The reform of local government debt management has weakened the government's implicit guarantee and administrative intervention, thereby enhancing the effectiveness of credit enhancement for city investment bonds. |
|---|---|---|---|
| 8 | 《Opinions on Properly Resolving the Follow-up Financing Issues of Under-construction Projects of Government-funded Platform Companies》 | 2015.05 | Relax the financing restrictions on banking institutions for city investment companies, providing conditions for dealing with the existing city investment bonds. |
| 9 | 《Notice on Issuing Emergency Disposal Plan for Local Government Debt Risks》 | 2016.10 | The regulation aims to break investors' implicit guarantee and faith in implicit guarantees, and sets a maximum compensation limit for government's outstanding guaranteed debt. |
| 10 | 《Notice on Further Standardizing the Borrowing and Financing Behavior of Local Governments》 | 2017.05 | Local governments and their departments shall not provide any form of guarantee for the debts of any units or individuals, and for the first time, it proposes to hold accountable, including local governments, enterprises, financial institutions, and other relevant responsible parties. |
| 11 | 《Notice on Strengthening the Ability of Corporate Bonds to Serve the Real Economy and Strictly Preventing Local Government Debt Risks》 | 2018.02 | City investment enterprises should proactively declare that they do not assume government financing functions and that the issuance of bonds does not involve the incurrence of new local government debt. |
| 12 | 《Notice on the Increase of Hidden Debt of Local Governments through PPP Projects》 | 2019.06 | Allow the hidden local government debt of city investment companies to be handled through debt restructuring. |
| 13 | 《Notice on Accelerating the Issuance and Use of Local Government Special Bonds》 | 2020.12 | Regulate local government debt by dividing it into "red, orange, yellow, and green" categories to prevent tail risks from municipal investment bonds. |
| 14 | 《Opinions on Further Deepening the Reform of the Budget Management System》 | 2021.04 | Emphasizing the importance of resolving hidden debt of local governments, placing greater emphasis on risk prevention, cleaning up and standardizing local financing platforms, and separating their government financing functions. |
| 15 | 《Guidelines for Further Promoting the Reform of Provincial and Lower-level | 2022.06 | Hedging against substantial default risks reflects the continuity of the central government's policy direction of |

|    | Financial Systems》 |  | "controlling the increase in hidden debt and stabilizing the existing debt" towards local governments. |
|----|----|----|----|
| 16 | 《Notice on Strengthening the Management of Government Investment Projects in Key Provinces (for Trial Implementation)》 | 2023.12 | It is required that key provinces strictly control the construction of new government investment projects until the local government debt risk is lowered to a medium or low level. |
| 17 | 《Notice on Further Coordinating the Prevention and Resolution of Local Government Debt Risks》 | 2024.03 | The 19 provinces are allowed to select their own areas with heavy debt burdens and high debt relief difficulties within their jurisdiction, with prefecture-level cities as the main focus. |

## 3.2 Keyword Statistical Analysis

Keyword statistical analysis is the first step of textual analysis of the policy documents on implicit government guarantees for municipal investment bonds. In this paper, keyword statistical analysis had been done by the Jieba library in Python program. We also conducted a comprehensive and systematic analysis of all policy documents related to municipal investment bonds and local government debt, and built a dictionary of specialized vocabulary for improving the accuracy and professionalism of the segmentation. The word frequency statistics of the standardized policy documents on implicit government guarantee of municipal investment bonds are showed in Table 5.

Table 5．Word Frequency Analysis of Policy Documents

| Keyword | Frequency | Keyword | Frequency | Keyword | Frequency |
|---|---|---|---|---|---|
| government | 668 | mechanism | 101 | Banking industry | 39 |
| Debt | 477 | Guarantee | 101 | Major | 38 |
| Place | 411 | Regulations | 100 | Regulation | 38 |
| Project | 321 | According to the law. | 99 | Fund | 38 |
| Fundraising | 314 | Investment | 97 | System | 38 |
| Managemen t | 298 | Standard | 97 | Comprehensive | 38 |
| Risk | 289 | Construction | 95 | Unity | 37 |
| platform | 216 | provincial | 91 | Newly added | 37 |
| Department | 195 | Bank | 88 | Law | 37 |
| Funds | 177 | Income | 86 | Effective | 47 |
| Loan | 173 | Implement | 86 | Monitoring | 36 |
| Bond | 165 | Disposal | 84 | Combine | 36 |
| budget | 164 | Policy | 83 | Market | 36 |
| fiscal | 232 | Emergency | 81 | Rectification | 36 |
| Business | 136 | Credit | 76 | Make sure | 35 |
| institution | 133 | Strict | 75 | Approval | 34 |
| Development | 130 | Society | 71 | Plan | 34 |
| Finance | 129 | Assets | 69 | Carry out | 34 |
| Expenditure | 121 | Violation | 68 | State-owned | 34 |
| governmental | 103 | Borrow money | 67 | Program | 34 |
| Perfection | 67 | Nation | 62 | Leadership | 59 |
| Reform | 67 | Security | 61 | Take on | 58 |
| Service | 66 | Capital | 61 | key point | 55 |
| Specialized | 64 | Public | 60 | Arrange | 53 |
| Payment | 63 | Principle | 59 | In time | 53 |
| Incorporate | 51 | In practice | 47 | Condition | 41 |
| Pay back | 50 | Prevention | 46 | economy | 41 |
| Pay off debt | 50 | Breaking the law | 46 | Organization | 41 |
| Supervision | 49 | Size | 45 | Land | 40 |
| problem | 49 | Existing stock | 44 | Check | 39 |
| people | 49 | Reasonable | 44 | ability | 39 |
| Evaluation | 48 | Measures | 44 | Adjustment | 42 |
| Transfer | 48 | Healthy | 42 | Foundation | 42 |

According to Table 5, it can be seen that "municipal investment bonds" and " implicit government guarantee" are the core words in the policy of municipal investment bonds. In addition to these, the policy of issuing municipal investment bonds also involves fiscal, risk, regulatory and market-related content, covering infrastructure and public service levels, with local governments and financing platforms being the key focus. Furthermore, the policy can be seen as focusing on standardizing the issuance and use of municipal investment bonds, ensuring the healthy development of the municipal investment bond market by regulating the financing behavior of local governments and controlling the debt risks of city investment enterprises, while maintaining the stability of the financial market and promoting sustained stable economic growth.

## 4 Calculation of PMC Index and Implicit Government Guarantee Strength

### 4.1 Descriptive statistics of Calculation Results

Based on the textual analysis of the 17 files, we calculated the main-variables、PMC index and implicit government guarantee strength. The descriptive statistical results are shown in Table 6.

According to Table 6, it can be showed that the main-variables have the following characteristics:

Table 6　Descriptive statistics of main-variables and PMC index

| Variables | Number | Mean | Standard deviation | Minimum value | Maximum value |
|---|---|---|---|---|---|
| $P_1$ | 17 | 0.51 | 0.20 | 0.17 | 0.83 |
| $P_2$ | 17 | 0.86 | 0.17 | 0.67 | 1.00 |
| $P_3$ | 17 | 0.39 | 0.21 | 0.00 | 0.60 |
| $P_4$ | 17 | 0.87 | 0.13 | 0.75 | 1.00 |
| $P_5$ | 17 | 0.34 | 0.20 | 0.00 | 0.60 |
| $P_6$ | 17 | 0.64 | 0.20 | 0.40 | 1.00 |
| $P_7$ | 17 | 0.52 | 0.22 | 0.20 | 1.00 |
| $P_8$ | 17 | 0.51 | 0.19 | 0.25 | 0.75 |
| $P_9$ | 17 | 0.75 | 0.09 | 0.60 | 0.80 |
| $P_{10}$ | 17 | 1.00 | 0.00 | 1.00 | 1.00 |
| PMC | 17 | 6.39 | 0.97 | 4.80 | 7.77 |
| G | 17 | 3.61 | 0.97 | 2.23 | 5.20 |

(1) Except for transparency of policy($P_{10}$), the variable with the highest mean is the policy sources ($P_4$), with an average value of 0.87, indicating that policies to regulate implicit government guarantees for municipal investment bond issuance are usually issued by higher-level institutions, which have significant impacts on national development, regional economies, and industrial structures.

(2) The lowest mean among the first-level variables is the areas of policy application ($P_3$), with an average value of 0.39, indicating that the policy of municipal investment bond has relatively low coverage in specific areas, which may limit the implementation effect of the policy.

(3) The average scores for validity of policy ($P_2$), incentives and guarantees of policy ($P_5$), and subjects addressed of policy ($P_8$) are all below 0.60, indicating that there is a lot of improvement for improvement in the diversity of policy impact objects regarding the implicit government guarantee of municipal investment bonds.

## 4.2 Trend of Implicit Government Guarantee Strength

Table 7 shows the trend of PMC index calculated between 2008-2024, which can be used to analyze the evolution trend of policy consistency and effect from the perspective of time change.

Table 7. Main-variables、PMC index and Implicit Government Guarantee

|     | 2008 | 2009 | 2010 | 2011 | 2012 | 2013 | 2014 | 2015 | 2016 | 2017 | 2018 | 2019 | 2020 | 2021 | 2022 | 2023 | 2024 |
|-----|------|------|------|------|------|------|------|------|------|------|------|------|------|------|------|------|------|
| P1  | 0.33 | 0.50 | 0.17 | 0.17 | 0.33 | 0.33 | 0.50 | 0.67 | 0.67 | 0.67 | 0.67 | 0.33 | 0.50 | 0.83 | 0.67 | 0.67 | 0.67 |
| P2  | 1.00 | 1.00 | 1.00 | 0.67 | 0.67 | 1.00 | 1.00 | 0.67 | 1.00 | 1.00 | 0.67 | 0.67 | 1.00 | 1.00 | 1.00 | 0.67 | 0.67 |
| P3  | 0.20 | 0.40 | 0.20 | 0.20 | 0.20 | 0.20 | 0.60 | 0.60 | 0.60 | 0.00 | 0.20 | 0.40 | 0.40 | 0.60 | 0.60 | 0.60 | 0.60 |
| P4  | 0.75 | 0.75 | 1.00 | 0.75 | 0.75 | 0.75 | 1.00 | 1.00 | 0.75 | 0.75 | 0.75 | 1.00 | 0.75 | 1.00 | 1.00 | 1.00 | 1.00 |
| P5  | 0.20 | 0.40 | 0.00 | 0.20 | 0.20 | 0.20 | 0.20 | 0.60 | 0.60 | 0.00 | 0.40 | 0.60 | 0.60 | 0.40 | 0.40 | 0.40 | 0.40 |
| P6  | 0.60 | 0.40 | 0.60 | 0.40 | 0.40 | 0.40 | 0.40 | 0.80 | 0.80 | 1.00 | 1.00 | 0.60 | 0.60 | 0.80 | 0.60 | 0.60 | 0.80 |
| P7  | 0.40 | 0.40 | 0.20 | 0.40 | 0.40 | 0.40 | 0.40 | 0.60 | 0.80 | 1.00 | 0.40 | 0.60 | 0.60 | 0.40 | 0.80 | 0.20 | 0.80 |
| P8  | 0.25 | 0.50 | 0.50 | 0.50 | 0.25 | 0.25 | 0.25 | 0.75 | 0.75 | 0.75 | 0.50 | 0.50 | 0.50 | 0.75 | 0.50 | 0.75 | 0.50 |
| P9  | 0.60 | 0.80 | 0.80 | 0.60 | 0.60 | 0.60 | 0.80 | 0.80 | 0.80 | 0.80 | 0.80 | 0.80 | 0.80 | 0.80 | 0.80 | 0.80 | 0.80 |
| P10 | 1.00 | 1.00 | 1.00 | 1.00 | 1.00 | 1.00 | 1.00 | 1.00 | 1.00 | 1.00 | 1.00 | 1.00 | 1.00 | 1.00 | 1.00 | 1.00 | 1.00 |
| PMC | 5.33 | 6.15 | 5.47 | 4.88 | 4.80 | 5.13 | 6.15 | 7.48 | 7.77 | 6.97 | 6.38 | 6.50 | 6.75 | 7.58 | 7.37 | 6.68 | 7.23 |
| $G$ | 4.67 | 3.85 | 4.53 | 5.12 | 5.20 | 4.87 | 3.85 | 2.52 | 2.23 | 3.03 | 3.62 | 3.50 | 3.25 | 2.42 | 2.63 | 3.32 | 2.77 |

From Table 7, it can be seen that the trend of the PMC index, which regulates the implicit government guarantee policy, exhibits the following characteristics:

(1) Validity of policy ($P_2$) reflects the timeliness and responsiveness of policy to market changes. From the data, the $P_2$ index reached its peak of 1.00 in 2008 and 2009, indicating that the government quickly acted during the financial crisis to ensure the timeliness of policy. However, it dropped to 0.67 in 2011, reflecting the gradual stabilization of the market and the reduced need for policy adjustment. In 2016 and beyond, the $P_2$ stabilized at 1.00 again, indicating that policy makers continued to pay attention to market dynamics and maintained the timeliness of policy.

(2) Function of policy ($P_6$) measures the effectiveness of policies in achieving predefined objectives. From the data, the P6 index increased from 0.60 in 2008 to 1.00 in 2016 and beyond, showing that the policy function has significantly improved over time, with policies becoming more effective in achieving their predefined objectives.

(3) Subjects addressed of policy ($P_8$) reflects the breadth and depth of policy impact. The $P_8$ increased from 0.25 in 2008 to 0.75 in 2016 and beyond, indicating that the policy's scope of influence has expanded to include more market participants and areas.

By analyzing above key variables, it is evident that these changes indicate that policy makers are constantly optimizing policies to adapt to market changes and needs, thereby enhancing the effectiveness for reducing the implicit government guarantee expectations and strength.

## 4.3 Trend of implicit government guarantee intensity

According to Table 6, The mean of the implicit government guarantee intensity(G) was determined to be 3.61, with a minimum of 2.23 and a maximum of 5.20, alongside a standard deviation of 0.97. This suggests that there remains potential for enhancing policies regarding the de-implicit government guarantee expectations for municipal investment bonds. specifically, the expectation surrounding implicit government guarantees persists at a relatively elevated level. Overall, the trend indicates that the implicit government guarantee intensity has fluctuated from 4.67 in 2008 to 2.77 in 2024, exhibiting a significant decline since 2015—an indication that policy effectiveness and intensity have markedly improved over time. In summary, while there is an overall downward trajectory in the intensity index, this reflects the gradual impact of governmental efforts aimed at regulating implicit guarantee policies.

Before 2013, the implicit government guarantee strength was relatively high, indicating that during this period, the government's risk management of municipal investment bonds was relatively weak, and indicating that the early policies were not strong enough, and the policy effects had not been fully manifested, resulting in a relatively large implicit government guarantee. After 2014, the implicit government guarantee strength decreased and stabilized at a lower level, indicating that the effects of the government's policy to regulate implicit guarantees for municipal investment bonds are gradually being realized, and the government's potential implicit guarantee is reduced.

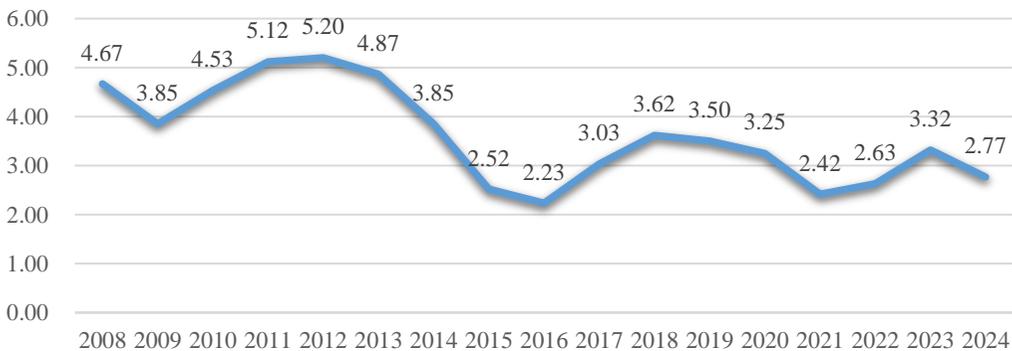

Figure 1. the trend of implicit government guarantee strength.

## 4.4 PMC Surface of implicit government guarantees strength

Based on the PMC index model, PMC surfaces were constructed (See Figure 2-5), and the spider web diagrams (Figure 6) were constructed by combining the input-output table and the government implicit guarantee index (G), which provide a clear analysis of the overall changes in the levels of implicit government guarantee based on the PMC index.

According to the PMC surface, 17 policies stand out in terms of policy effectiveness, issuing authority, and legal level, indicating that these policies can respond quickly to market changes, are issued by authoritative institutions, and have high legal effectiveness. However, these policies have shortcomings in terms of clarity of policy nature, coverage of policy fields, and incentive and guarantee measures, which may be key factors affecting the overall removal of government implicit guarantees.

As shown in Figure 2, the policy on implicit guarantee for municipal investment bonds scored high in terms of policy timeliness, indicating that the policy could adapt quickly to market changes, indicating that regulatory authorities were able to identify the existence of market expectations for implicit guarantee of municipal investment bonds in a timely manner and also that they found that such implicit guarantee expectations were not good for the risk management of municipal investment bonds, so they promptly issued relevant policies. However, the scores of all policies in terms of clarity of policy nature were low. This may mean that the policy objectives and implementation standards are not clear enough, leading to uncertainty in market expectations and responses to the policy. If the policy has higher clarity, it may enhance market confidence and stability, thereby reducing the government's implicit guarantee. At the same time, it was also found that if the policy could provide reasonable incentives and guarantees, it may more effectively promote the active participation of market participants and reduce potential moral hazard, thereby reducing the government's implicit guarantee.

From Figure 3 policies with extremely strong implicit guarantees, although they score low in terms of policy domain and incentive assurance, perform well in terms of policy timeliness and the authority of the issuing institution, indicating that these policies can be timely issued by authoritative institutions, but need to be strengthened in terms of coverage in specific areas and incentives for market participants.

From Figure 4, it can be seen that policies with stronger implicit guarantees generally score higher in most indicators, especially in terms of policy timeliness and the authority of the issuing institution, but also have shortcomings in terms of policy nature and incentive assurance, indicating that although these policies perform well in certain key areas, improvements are still needed in terms of clarity and incentives.

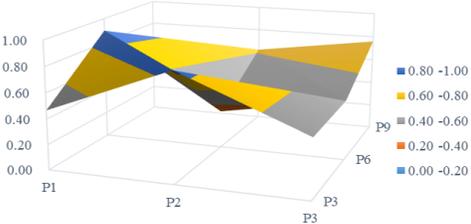

Figure 2．PMC surfaces of 17 policies

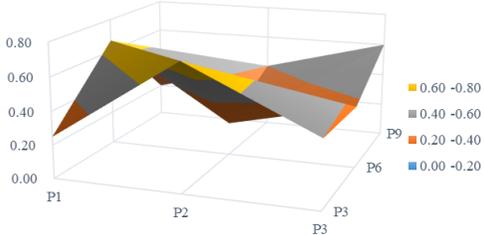

Figure 3. PMC surfaces with Very strong implicit government guarantee

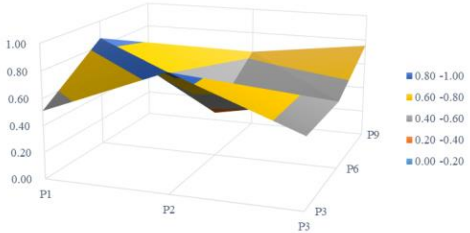

Figure 4. PMC surfaces with stronger implicit government guarantee

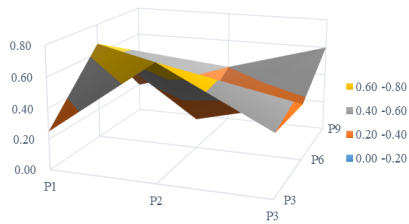

Figure 5. PMC surfaces with lower implicit government guarantee

As shown in Figure 5, policies with lower implicit government guarantee scored lower in terms of policy effectiveness, policy function, and impact at different levels, indicating that although these policies are issued by authoritative institutions with high legal validity, they need to be further improved in terms of timeliness, comprehensiveness of functions, and impact at different levels.

It is worth noting that policies with extremely strong guarantee intensity and policies with stronger guarantee intensity may neglect guarantee intensity in some aspects, which may lead to an underestimation of market risks and increase moral risks. Therefore, these policies need to be further improved in terms of timeliness, comprehensiveness, and their impact at different levels to achieve a more balanced and sustainable market development.

Overall, whether it is a policy with very strong guarantees or a policy with stronger guarantees, or a policy with lower guarantees, all need to be strengthened in terms of the clarity of policy nature, the coverage of policy fields, and the incentive and guarantee measures. In particular, the incentive and guarantee measures seem to be a common weakness that affects the strength of all policy guarantees. By improving these aspects, we can reduce market expectations of implicit guarantees and thus promote market stability and investor confidence more effectively.

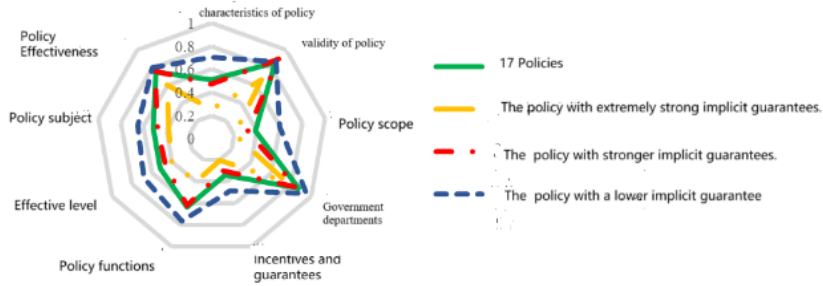

Figure 6. The spider web diagrams

In the current economic and financial context of China, the government's implicit guarantee policy for city investment bonds is crucial to the stability of the market and investor confidence. A comprehensive analysis of the spider web diagram shows that lowering the government's implicit guarantee strength, combined with clarity, timeliness, incentive assurance, coverage scope, and risk control, is the key to achieving stability and healthy development of the city investment bond market. Policy makers need to consider and balance these aspects to ensure that the policy can effectively support the market without triggering excessive dependence and risk accumulation.

## 5 Conclusions

At certain times, the implicit government guarantee of state-owned enterprise bonds is not just a matter of expectation. There have been many real cases where the Chinese government has provided support for bonds issued by state-owned enterprises. However, the implicit government guarantee is not conducive to the recognition and management of risks by all parties in the bond market, and it will also lead to excessive debt of local governments and state-owned enterprises. Therefore, various departments of the central government have issued a series of policies documents, aiming to strictly manage the risks of municipal investment bonds and break the expectation of implicit government guarantees and repayment.

This paper uses textual analysis methods and the PMC index to evaluate the effectiveness of these policies, and proposes a new calculation method for the implicit government guarantee intensity. It is opposite to the evaluation of the effectiveness of policy de-implicit guarantee, if the effect of policy de-implicit guarantee is better, the implicit government guarantee intensity will be lower, the opposite is also true. From the research findings, the introduction of a series of policies has indeed reduced the implicit government guarantee, indicating that the risk management policies for municipal investment bonds adopted by various levels of government are effective.

Our study also found that the changes in the intensity of implicit government guarantees come from different aspects of policy. For example, the government at all levels and departments scored high in the timeliness of policy formulation for municipal investment bonds, which indicates that they can quickly adapt to market changes and make timely risk control measures to address the implicit guarantee expectations and development. In the clarity of policy nature, the PMC evaluation score was lower, which may mean that the policy objectives and implementation standards are not clear enough, leading to uncertainty in the market's expectations and responses to the policy. Furthermore, the policy incentive and guarantee system are not good enough, which leads to the need to further improve the expectation of removing implicit government guarantees.

In summary, this paper uses text mining techniques to analyze the 17 policy documents related to municipal investment bonds in the past decade, and proposes a new method for measuring the strength of implicit government guarantee. It also provides an objective and scientific evaluation tool for investors and regulatory authorities. Of course, there may still be a lot of work for improvement in the precise identification of keywords in the text analysis process and the various indicators of the PMC policy rating.